\pdfoutput=1
\RequirePackage{ifpdf}
\ifpdf
\documentclass[pdftex]{sigma}
\else
\documentclass{sigma}
\fi

\numberwithin{equation}{section}

\begin{document}
\allowdisplaybreaks

\renewcommand{\thefootnote}{$\star$}

\renewcommand{\PaperNumber}{004}

\FirstPageHeading

\ShortArticleName{Dynamical Equations of Mechanical Systems with Position-Dependent Mass}

\ArticleName{Dynamical Equations, Invariants and Spectrum\\ Generating Algebras of Mechanical
Systems\\ with Position-Dependent Mass\footnote{This
paper is a~contribution to the Special Issue ``Superintegrability, Exact Solvability, and Special
Functions''.
The full collection is available at
\href{http://www.emis.de/journals/SIGMA/SESSF2012.html}{http://www.emis.de/journals/SIGMA/SESSF2012.html}}}

\Author{Sara CRUZ~Y~CRUZ~${}^\dag$ and Oscar ROSAS-ORTIZ~${}^\ddag$}

\AuthorNameForHeading{S.~Cruz~y~Cruz and O.~Rosas-Ortiz}

\Address{${}^\dag$~SEPI-UPIITA, Instituto Polit\'ecnico Nacional, Av. IPN No. 2580,\\
\hphantom{${}^\dag$}~Col.
La Laguna Ticom\'an, C.P. 07340 M\'exico D.F.
Mexico}
\EmailD{\href{mailto:sgcruzc@ipn.mx}{sgcruzc@ipn.mx}}

\Address{${}^\ddag$~Physics Department, Cinvestav, A.P. 14740,
M\'exico D.F. 07000, Mexico}
\EmailD{\href{mailto:orosas@f\/is.cinvestav.mx}{orosas@f\/is.cinvestav.mx}}

\ArticleDates{Received July 31, 2012, in f\/inal form January 12, 2013; Published online January 17, 2013}

\Abstract{We analyze the dynamical equations obeyed by a~classical system with position-dependent mass.
It is shown that there is a~non-conservative force quadratic in the velocity associated to the
variable mass.
We construct the Lagrangian and the Hamiltonian for this system and f\/ind the modif\/ications required
in the Euler--Lagrange and Hamilton's equations to reproduce the appropriate Newton's dynamical law.
Since the Hamiltonian is not time invariant, we get a~constant of motion suited to write the
dynamical equations in the form of the Hamilton's ones.
The time-dependent f\/irst integrals of motion are then obtained from the factorization of such a
constant.
A canonical transformation is found to map the variable mass equations to those of a~constant mass.
As particular cases, we recover some recent results for which the dependence of the mass on the
position was already unnoticed, and f\/ind new solvable potentials of the P\"oschl--Teller form which
seem to be new.
The latter are associated to either the $su(1,1)$ or the $su(2)$ Lie algebras depending on the sign
of the Hamiltonian.}


\Classification{35Q99; 37J99; 70H03; 70H05}

\renewcommand{\thefootnote}{\arabic{footnote}}
\setcounter{footnote}{0}

\section{Introduction}

In recent papers~\cite{Cru08, Cru07,Cru08b,Kur08} the factorization of the classical Hamiltonian in
terms of two functions that, together with the Hamiltonian itself, lead to a~Poisson algebra was
discussed.
This program included mechanical systems with mass explicitly dependent on position
\cite{Cru07,Cru08b} and was extended to the quantum case~\cite{Cru07,Cru08b,Cru09,Cru10} (other
already reported approaches can be found in~\cite{Ald05,Bag05, Que04}).
Position-dependent mass functions~$m(x)$ give rise to `forces quadratic in the velocity' which, in
turn, lead to nonlinear dif\/ferential equations of motion in the Newtonian approach.
The most celebrated example of this kind of equations, due to Mathews and Lakshmanan~\cite{Mat74}
(see also~\cite{Car07, Car04}), corresponds to the nonlinear oscillator
\begin{gather}
\big(1+\lambda x^2\big) \ddot x - \lambda x \dot x^2 + \alpha^2 x=0,
\label{Ma1}
\end{gather}
and is derivable from the Lagrangian
\begin{gather*}
L= \frac12 \left( \frac{1}{1+ \lambda x^2} \right) \big(\dot x^2 -\alpha^2 x^2 \big).
\end{gather*}
Masses varying with the position can be also associated to the kinetic energy of dynamical systems
in curved spaces with either constant curvature~\cite{Car12, Koz92,Sla00}, or non-constant curvatu\-re~\cite{Bal11, Rag10}.
Similar relationships arise in geometric optics where the position dependent refractive index can
be interpreted as a~variable mass~\cite{Wol04}.
In semiconductor theory, it has been found that the coherent superpositions of states connected to
dif\/ferent masses are forbidden~\cite{von82}.
Moreover, there is not unicity in the construction of quantum Hamiltonians (see, e.g.,
\cite{Cru07,Cru09} and references quoted therein), and the invariance under the change of inertial
frames is not granted for position-dependent mass systems~\cite{von83} (however, see the discussion
on the matter in~\cite{Lev95}).
In general, the variable mass dynamics involves masses which are functions of either position or
time (or functions of time and position).
Besides the works mentioned above, the applications include the motion of rockets~\cite{Som64}, the
raindrop problem~\cite{Kra81}, the variable mass oscillator \cite{Flo03}, the inversion potential
for NH$_3$ in density theory~\cite{Aqu98}, the two-body problem associated to the evolution of
binary systems~\cite{Had67}, the ef\/fects of galactic mass loss \cite{Rich82}, neutrino mass
oscillations~\cite{Bet82}, and the problem of a~rigid body against a~liquid free surface
\cite{Pes03} among other.
Yet, the discussion is far from being exhausted.
For instance, it has been indicated that Newton's second law is valid only for constant mass and
that, if the mass variation is due to accretion or ablation, the corresponding equation must be
modif\/ied~\cite{Pla92}.
Other works report the conservation laws for a~variable mass system~\cite{Cve93} and the
construction of standard and non-standard Lagrangians for dissipative variable mass systems
\cite{Mus08}.

The present work deals with the dynamical equations obeyed by classical position-dependent mass systems.
We analyze the problem in the Newton, Lagrange and Hamilton approaches by deriving in each case the
equations of motion.
A non-conservative force which is quadratic in the velocity appears because of the mass-function
$m(x)$ and induces modif\/ications in the dynamical laws with respect to the constant mass case.
As a~consequence, the position-dependent mass Hamiltonian ${\cal H}=\frac{p^2}{2m(x)}+{\cal
V}(x)$ is not time-independent.
Hence, the force due to the variable mass can be associated to dissipation.
Despite this result, it is possible to construct a~constant of motion~$H$ (in energy units) that
allows to express the dynamical equations in the conventional Hamilton's form for the appropriate
phase space.
Using this invariant and the factorization method the problem can be solved in an algebraic form.

\looseness=-1
The organization of the paper is as follows.
In Section~\ref{deformed}, we depart from the Newton's equation of motion for these systems and
show that it is possible to get a~Lagrangian in the standard form $L=\frac{p^2}{2m(x)}-{\cal
V}(x)$, at the cost of modifying the Euler--Lagrange equations of motion.
The modif\/ication must include the non-inertial force which is quadratic in the velocity.
Then, the variable mass Hamiltonian ${\cal H}$ is constructed and the related equations of motion
are shown to be a~modif\/ied version of the conventional Hamilton's equations while they are
consistent with both, the Newtonian and the Lagrangian formulations already discussed.
It is shown that the time derivative of the Hamiltonian ${\cal H}$ is cubic in the velocity.
In Section~\ref{constant} we construct the constant of motion $H$ and show that the related
dynamical equations acquire the form of the conventional Hamilton's equations.
Although the Lagrangian and Hamiltonian are not unique, our expressions for~$L$, ${\cal H}$, and~$H$ are found to be in correspondence with the general expressions derived by other authors
\cite{Mus08}.
In Section~\ref{first}, some other results already reported for constant masses in presence of
forces quadratic in the velocity are shown to be particular cases of our approach for specif\/ic
nontrivial forms of the mass-function~$m(x)$.
In Section~\ref{secfac}, the factorization method discussed in~\cite{Cru08, Cru07,Cru08b,Kur08} is
extended to the case of the variable mass systems we deal with in this paper.
As it will be shown, the underlying Poisson structures lead to a~deformed Poisson algebra only if
the potential is of the P\"oschl--Teller form.
Then, explicit expressions for the phase trajectories described by these variable masses are found.
Section~\ref{point} is devoted to the analysis of a~point transformation that leads from the
variable mass problem to the one of a~constant mass.
This transformation is shown to be canonical in the sense that it leaves invariant the Hamilton's
equations of motion here derived.
In Section~\ref{PTeller} we analyze the diverse P\"oschl--Teller potentials we have at hand by
supplying specif\/ic mass-functions~$m(x)$ in our approach.
In particular, we use a~doubly singular mass $m_{\rm ss}(x)$ arising from the analysis of the inversion
potential for NH$_3$ in terms of the density operator~\cite{Aqu98}.
The potentials arising from the singular as well as from the regular masses which we have used in
our previous works are also analyzed.
Finally, an exponential mass-function~$m_{\rm e}(x)$, introduced here to recover the Lagrangian forms
reported by other authors for constant mass systems, gives rise to solvable potentials which, as
far as we know, have been not reported previously in the literature.
The paper is closed with some conclusions.

\section{Deformed algebras of a~particle with position dependent mass}\label{deformed}

\subsection{Newtonian framework}\label{new}

For the sake of motivation consider the one-dimensional motion of a~system of mass $m(x)>0$,
explicitly dependent on the position~$x$, acted by a~force~$F$ which in general depends on position~$x$,
velocity $\dot x$, and time $t$ (the case of~$n$ generalized coordinates is straightforward).
The Newton's equation of motion is
\begin{gather}
F(x, \dot x; t) = \frac{dp}{dt}= m'(x) \dot x^2 + m(x) \ddot x,
\label{force}
\end{gather}
where $p=m(x)\dot x$ is the linear momentum.
Hereafter $\dot g$ and $g'$ will stand for time and position derivatives of~$g$ respectively.
To get~\eqref{force} we have assumed a~null velocity~$v_m$ for the accreted or emitted mass
relative to the system (otherwise, one of the velocities in the quadratic term $m'(x)\dot x\dot
x$ must be substituted by the displacement $v_m+\dot x$).
Equation~\eqref{force} has the general form
\begin{gather*}
F(x, \dot x; t) = \alpha(x;t) \ddot x + \beta(x,\dot x; t),
\end{gather*}
and can be decoupled in the nonlinear system
\begin{gather}
\dot x =v, \qquad \dot v = \frac{1}{m(x)}\left[ F(x,v;t) - m'(x) v^2\right].
\label{non1}
\end{gather}
This def\/ines a~path on the plane $(x,v)$ in terms of the parameter $t$ (see, e.g.,~\cite{Arn78}).
The latter expressions acquire the standard form whenever $m'=0$:
\begin{gather}
\dot x =v, \qquad \dot v = \frac{1}{m_0} F_0 (x,v;t).
\label{non2}
\end{gather}
Here the constant $m_0>0$ corresponds to the mass of a~system subjected to the action of the force
$F_0$.
It is possible to make the equivalence between the two systems,~\eqref{non1} and \eqref{non2}, in
such a~way that the variable mass~$m(x)$ describes the same path as a~given constant mass $m_0$.
For instance, consider the mass
\begin{gather}
m(x) = -m_0 [1 - k V_0(x)]
\label{mLanc}
\end{gather}
with $k$ a~constant and $V_0(x)$ a~dif\/ferentiable function.
Let $E_0=1/k$ be the total energy of $m_0$, then equations~\eqref{non1} and~\eqref{non2} lead to equivalent paths provided that
\begin{gather*}
F_0 = -\frac{2E_0}{m_0 v^2} F + 2\frac{d V_0(x)}{dx}.
\end{gather*}
From this expression, notice in particular that the variable mass~\eqref{mLanc} describes, under
its own inertia (i.e., in absence of the external f\/ield~$F$), exactly the same path as the constant
mass $m_0$ under the action of the potential function $-2V_0(x)$.
This example shows that the mechanical energy $E_0$ of the properly chosen constant mass system can
be used to parameterize the path described by the variable mass~$m(x)$ on the $(x,v)$-plane
(regardless the mechanical energy is not conserved in the latter system).
However, we have to emphasize that this equivalence holds for the paths only, not for the motion
with respect to time~\cite{Lan93}.

To get some insight on the forces involved in the Newton's equation~\eqref{force} let us rewrite
this in the standard form
\begin{gather}
m(x) \ddot x = F_{\rm net}(x, \dot x; t) \equiv F(x, \dot x; t) - m'(x) \dot x^2.
\label{force2}
\end{gather}
Since $\dot x^2>0$, the system is accelerated if the rate $m'(x)$ is negative, and decelerated if
this is positive.
Thus, the term quadratic in the velocity $-m'(x)\dot x^2$ corresponds to the thrust of the system,
and~\eqref{force2} indicates how this alters the velocity $\dot x$.
In this way, the net force $F_{\rm net}$ on a~particle suf\/fering a~spatial variation of its mass
results as a~combination of the external force~$F$ and the thrust $-m'(x)\dot x^2$.

Mechanical systems of variable mass are mostly studied in the cases where the mass~$m$ is an
explicit function of time~$m(t)$.
Immediate applications involve rockets and jet engines analyzed in terms of Newtonian theory~\cite{Som64}.
The main dif\/f\/iculty is that not all the involved forces are derivable from an ordinary potential
function or even from a~generalized potential.
In many situations of interest, the action of such forces produces the non-conservation of the
mechanical energy (see for instance the analysis of the raindrop problem in~\cite{Kra81}).
Nonetheless, the dynamics of time-dependent mass systems can be studied in the Lagrangian approach.
Indeed, for these systems in one degree of freedom and forces independent on velocity, Darboux
showed that it is always possible to construct a~Lagrangian~\cite{Dar94} (see also~\cite{Leu90}).
The equation of motion for some of these systems is connected to that of a~pendulum whose length~$\ell$ varies with time in the small-angle approximation $\ell\ddot\theta+2\dot\ell\dot
\theta+g\theta=0$ ($0<\theta<<1$).
For instance, a~time-dependent mass (undamped) oscillator obeys the equation $m\ddot x+\dot m
\dot x+kx=0$.
Here, the counterpart of~$\theta$ is the displacement~$x$ for the oscillator and the thrust $-\dot
m\dot x$ simulates a~damping linear in the velocity.
It is well known that the solutions of this last equation are either decreasing or increasing
oscillatory functions, depending on the sign of~$\dot m$.
Interestingly, a~simple experimental setup can be used to show that the standard expression for the
oscillator's frequency $\omega=\sqrt{k/m}$ remains valid if the mass is time-dependent~\cite{Flo03}; a~useful result since the mathematical tools used to analyze the constant mass
oscillators can also be applied if the mass depends on time.
Our approach considers a~mass which is an explicit function of position~$m(x)$, so that implicit
dependence on time is assumed~$m(x(t))$.
According to~\eqref{force}, the equation of motion of a~position-dependent mass oscillator is
nonlinear $m(x)\ddot x+m'(x)\dot x^2+kx=0$ (compare with equation~\eqref{Ma1}).
In contraposition to the time-dependent mass case, the thrust $-m'(x)\dot x^2$ is here quadratic
in the velocity.
We stress that, in general, it is not evident how a~set of canonical variables can be found such
that the Newton's equation~\eqref{force} is fulf\/illed.

In the sequel we shall introduce a~mechanism to determine the phase trajectories $(x(t),p(t))$
associated to the Newton's equation of motion~\eqref{force} for interactions~$F$ derivable from a
properly chosen potential function.
We will focus on the invariants which arise after factorizing the related Hamiltonian.
With this aim, we f\/irst analyze the problem in the Lagrangian and Hamiltonian frameworks to
determine whether the standard expression of the Hamiltonian ${\cal H}=\frac{p^2}{2m}+{\cal V}$
remains valid if the mass depends on position.

\subsection{Canonical framework}

Departing from the Newton's equation of motion~\eqref{force} for a~force independent on velocity,
and applying the D'Alembert's principle we arrive at
\begin{gather}
\frac{d}{dt} \left( \frac{\partial T}{\partial \dot x} \right) - \frac{\partial T}{\partial x} =
F(x; t) + \widetilde R,
\label{lag1}
\end{gather}
with $T$ and $\widetilde R$ the kinetic energy and reacting thrust, respectively given by
\begin{gather*}
T:= \frac12 m(x) \dot x^2, \qquad \widetilde R(x,\dot x; t): = -\frac12 m'(x) \dot x^2.
\end{gather*}
When the external force is derivable from a~scalar potential function $F=-\frac{\partial{\cal
V}}{\partial x}$, and this last does not depend on either velocity or time, equation~\eqref{lag1}
becomes
\begin{gather}
\frac{d}{dt} \left( \frac{\partial L}{\partial \dot x} \right) - \frac{\partial L}{\partial x} =
\widetilde R, \qquad L=T -{\cal V}.
\label{lag2}
\end{gather}
This last is the Lagrangian form of the Newton's equation of motion~\eqref{force}.
To verify that the Lagrange equation~\eqref{lag2} encodes the dynamics of the variable mass~$m(x)$
we are dealing with, consider that the acceleration $a=\ddot x$ arising from $L$ satisf\/ies
\begin{gather}
M a = Q,
\label{hess}
\end{gather}
with $M=\frac{\partial^2L}{\partial^2\dot x}$ the Hessian (matrix) of $L$ with respect to the
velocity, and the ``Lagrangian force'' $Q$ def\/ined as
\begin{gather}
Q = \frac{\partial L}{\partial x} - \dot x \frac{\partial^2 L}{\partial x \partial \dot x} +
\widetilde R.
\label{hess2}
\end{gather}
The force $Q$ includes the applied (physical) force $-\frac{\partial V}{\partial x}$, and the
f\/ictitious (non-inertial) force $\frac{\partial T}{\partial x}-\dot x\frac{\partial^2
T}{\partial x\partial\dot x}=-\widetilde R$.
This last shows that the reacting thrust $\widetilde R$ is a~non-inertial force.
The substitution of $L=\frac12m(x)\dot x^2-{\cal V}(x)$ in~\eqref{hess} and~\eqref{hess2} reproduces
the dynamical law~\eqref{force}.
Moreover $M=m(x)>0$, so that the Lagrangian $L$ is regular and equation~\eqref{hess} can be solved
for the acceleration $a=Q/m(x)$.
The solutions~$x(t)$ of the second order dif\/ferential equation $\ddot x=a(x,\dot x;t)$ can be
obtained from the nonlinear system~\eqref{non1}.

Once we have constructed the Lagrangian $L$ for the position-dependent mass~$m(x)$, we use the
momentum $p=\frac{\partial L}{\partial\dot x}=m(x)\dot x$ to obtain the Hamiltonian ${\cal H}$
from the Legendre transformation
\begin{gather}
{\cal H}(x,p;t) = p\dot x -L(x,\dot x; t) = \frac{p^2}{2m(x)} + {\cal V}(x).
\label{legendre}
\end{gather}
The Hamiltonian's time rate of change
\begin{gather}
\frac{d}{dt} {\cal H} = \widetilde R \dot x = -\frac12 m'(x) \dot x^3,
\label{power}
\end{gather}
shows that the value ${\cal E}$ of ${\cal H}$ is not independent of~$t$.
Since $\dot{\cal H}$ is cubic in the velocity, equation~\eqref{power} makes plain that the
variable mass system is dissipative~\cite{Mus08} (in the quantum case, the factorization method
applied to dissipative systems is associated to complex Riccati equations~\cite{Cas13}).
The time derivative~\eqref{power} can be also expressed in canonical form $\dot{\cal H}=
\frac{p}{m(x)}R$, with the reacting thrust $R$ rewritten as
\begin{gather}
R = R(x,p;t) = \widetilde R(x, \dot x(x,p; t); t) = -\frac{m'(x)}{m(x)} \left( \frac{p^2}{2m(x)}
\right).
\label{Rp}
\end{gather}
On the other hand, a~simple calculation leads to the phase space form of equation~\eqref{lag2}.
We get
\begin{gather}
\dot x = \frac{\partial {\cal H}}{\partial p}, \qquad \dot p =- \frac{\partial {\cal H}}{\partial x}+R.
\label{canonical}
\end{gather}
The dif\/ferential equations~\eqref{canonical} correspond to the law of motion of the mechanical
position-dependent mass system in phase space.
At time $t$, their solutions are the canonically conjugate variables of position and momentum $\{
x,p\}=1$, with $\{\,,\,\}$ the Poisson bracket
\begin{gather*}
\{f, g\} = \frac{\partial f}{\partial x} \frac{\partial g}{\partial p} - \frac{\partial f}{\partial
p} \frac{\partial g}{\partial x}.
\end{gather*}
Hence, the Hamilton's equations~\eqref{canonical} in the Poisson bracket formulation read
\begin{gather}
\dot x = \{ x, {\cal H} \}, \qquad \dot p = \{ p, {\cal H} \} +R.
\label{canonicalp}
\end{gather}
In general, the explicit form of $m(x)$ and ${\cal V}(x)$ determine the Hamiltonian's domain of
def\/ini\-tion~${\cal D}({\cal H})$.
A point transformation is then expected to reduce either equations~\eqref{canonical} or
\eqref{canonicalp} to the ones associated with an equivalent constant mass system in the
appropriate domain (see Section~\ref{point}).

To summarize our f\/irst results, let us rewrite the Lagrangian and the Hamiltonian associated to the
Newton's equation of motion~\eqref{force}, we have
\begin{gather}
L= \frac12 m(x) \dot x^2 - {\cal V}(x), \qquad {\cal H} = \frac{p^2}{2m(x)} + {\cal V}(x).
\label{LH}
\end{gather}
These results show that the standard expressions for $L$ and ${\cal H}$ remain valid if the mass is
an explicit function of the position.
Thus, a~practical description of the position-dependent mass particle interacting with an external
environment consists of replacing the (constant) mass $m_0$ by the appropriate function of the
position $m(x)$ in the conventional expressions $L_0=\frac{m_0}{2}\dot x^2-{\cal V}(x)$
and $H_0=\frac{p^2}{2m_0}+{\cal V}(x)$.
The price to pay is, however, that the Euler--Lagrange and the Hamilton's equations are not
expressed in the standard form.

\subsubsection{Energy constant of motion}\label{constant}

As we have seen, the systems with position-dependent mass are describable by either a~Lagrangian or
a Hamiltonian function.
However, the Hamiltonian ${\cal H}=\frac{p^2}{2m(x)}+{\cal V}(x)$, obtained from the replacing
of the constant mass $m_0$ by a~function of the position $m(x)$ in $H_0=\frac{p^2}{2m_0}+{\cal
V}(x)$, is not time-independent (see equation~\eqref{power}).
In this section we are going to construct a~constant of motion for such a~system.
As the equation~\eqref{canonical} is equivalent to the nonlinear system \eqref{non1}, we f\/irst look
for an invariant in the $(x,v)$-plane representation.
For our variable mass system, a~function~$I$ of~$x$ and~$v$ is a~constant of motion whenever the
following equation is true
\begin{gather*}
\frac{d}{dt} I = \dot x \frac{\partial I}{\partial x} + \dot v \frac{\partial I}{\partial v} = v
\frac{\partial I}{\partial x} + \left[ \frac{F}{m} -\frac{m'}{m} v^2 \right] \frac{\partial
I}{\partial v} =0.
\end{gather*}
Here we have used~\eqref{non1}.
After multiplying by $\phi(x,v)$, a~function on $(x,v)$ to be determined, one gets
\begin{gather}
\frac{\partial I}{\partial x} = \left[ \phi \frac{m'}{m} \right] v^2 - \frac{F}{m} \phi,
\label{inv2a}\\
\frac{\partial I}{\partial v} = v\phi.
\label{inv2b}
\end{gather}
Note that $\phi=\alpha m^{\beta}$, with $\alpha$ and $\beta$ constants, allows to write $\phi
\frac{m'}{m}=\frac{1}{\beta}\frac{\partial\phi}{\partial x}$.
Therefore~\eqref{inv2a} can be expressed as
\begin{gather}
\frac{\partial I}{\partial x} = \frac{v^2}{\beta} \frac{\partial \phi}{\partial x} -\frac{F}{m}
\phi = \frac{\partial}{\partial x} \left[ \frac{v^2}{\beta} \phi - \int^x \frac{F}{m} \phi dr
\right].
\label{inv3}
\end{gather}
The introduction of~\eqref{inv3} in \eqref{inv2b} gives $\frac{\partial I}{\partial v}=
\frac{2}{\beta}v\phi$.
Then we make $\beta=2$ to get
\begin{gather}
I = \alpha \left[ \frac{m^2 v^2}{2} + \int^x m \left( \frac{\partial \cal V}{\partial r} \right) dr
\right].
\label{inv4}
\end{gather}
Taking $\alpha=m_0^{-1}$ with $m_0$ in mass units we see that~\eqref{inv4} is a~constant of
motion expressed in energy units, we write
\begin{gather}
{\cal H}_{(x,v)} =\frac{m^2 v^2}{2 m_0} + \int^x \frac{m}{m_0} \left( \frac{\partial \cal
V}{\partial r} \right) dr.
\label{inv5}
\end{gather}
Thus, the invariant ${\cal H}_{(x,v)}$ represents the conservative Hamiltonian of our variable mass
system in the $(x,v)$-plane.
(Note even that this is reduced to a~constant mass Hamiltonian in the case $m=m_0$.)
The statement is ref\/ined using integration by parts
\begin{gather}
{\cal H}_{(x,v)} = \frac{m}{m_0} \widetilde{\cal H} \equiv \frac{m}{m_0} \left[ \frac{p^2}{2m} +
\widetilde{\cal V} \right],
\label{inv6}
\end{gather}
where
\begin{gather*}
\widetilde{\cal V} = {\cal V} - \frac{m_0}{m} \int^x \left( \frac{m}{m_0} \right)' {\cal V} dr
\end{gather*}
is a~modif\/ication of the potential ${\cal V}$ in the phase space representation.
As we can see, the Hamiltonians ${\cal H}_{(x,v)}$ and $\widetilde{\cal H}$ are equivalent up to
the multiplicative function $\frac{m}{m_0}$.
Clearly $\widetilde{\cal H}$ is the variable mass Hamiltonian ${\cal H}$ plus a~function of the
position def\/ined by the potential ${\cal V}$.
As ${\cal H}_{(x,v)}$ is a~constant of motion, equation~\eqref{inv6} makes plain what is missing in
${\cal H}$ to be time-independent.
Moreover, solving~\eqref{inv6} for ${\cal H}$ yields
\begin{gather*}
{\cal H}=\frac{m_0}{m}\left[{\cal H}_{(x,v)}+\int^x\left(\frac{m}{m_0}\right)'{\cal V}dr\right]
=\frac{p^2}{2m}+{\cal V},
\end{gather*}
which is consistent with the def\/intion of ${\cal H}$ given in~\eqref{legendre}.

Let us take full advantage of the invariant~\eqref{inv5}.
For this, note that the f\/irst additive term in~\eqref{inv5} can be expressed in momentum-like form
by using
\begin{gather*}
\mathrm p:= \mathrm m v, \qquad \mathrm m:= \frac{m^2}{m_0}.
\end{gather*}
Therefore
\begin{gather}
H = \frac{\mathrm p^2}{2 \mathrm m} + {\cal V}_{\text{ef\/f}}(x), \qquad {\cal V}_{\text{ef\/f}}(x) =
\int^x\frac{m}{m_0} \left(\frac{\partial \cal V}{\partial r}\right) dr,
\label{hg}
\end{gather}
is the Hamiltonian ${\cal H}_{(x,v)}$ in the $(x,{\mathrm p})$-plane representation.
Concerning the equations of motion, the straightforward calculation leads to
\begin{gather}
\dot x = \frac{\partial H}{\partial {\mathrm p}}, \qquad -\dot{\mathrm p} = \frac{\partial H}{\partial x},
\label{inv6b}
\end{gather}
where we have used~\eqref{canonicalp}, \eqref{Rp} and \eqref{non1}.
Then, the time-variation of $f$, an arbitrary function of~$x$, $\mathrm p$, and $t$, is ruled by
\begin{gather}
\frac{d}{dt} f = \left\{ f, H \right\}_{x, \mathrm{p}} + \frac{\partial f}{\partial t},
\label{inv7}
\end{gather}
where
\begin{gather}
\left\{ f, g \right\}_{x, \mathrm p} = \frac{\partial f}{\partial x} \frac{\partial g}{\partial
\mathrm p} -\frac{\partial f}{\partial \mathrm p} \frac{\partial g}{\partial x}.
\label{inv7a}
\end{gather}
In particular, if $f\!=\!H$ then $\frac{d}{dt}H\!=\!0$.
On the other hand, using~\eqref{inv7a} we realize that $\{x,\mathrm p\}_{x,\mathrm p}\!=\!1$, so
that~$x$ and $\mathrm p$ are conjugate variables.
The equations of motion~\eqref{inv6b} can be now expressed~as
\begin{gather}
\dot x = \{ x, H\}_{x, \mathrm p}, \qquad \dot{\mathrm p} = \{ \mathrm p, H\}_{x, \mathrm p}.
\label{inv6c}
\end{gather}
For the sake of simplicity, hereafter we shall omit the sublabels in the brackets $\{\cdot,\cdot\}_{x,\mathrm p}$.

\looseness=-1
At this stage some words on the dif\/ferences between the dynamics of the Hamiltonian ${\cal H}$ and
that of $H$ are necessary.
In the former case, for a~variable mass $m(x)$ subjected to a~given potential ${\cal V}(x)$, the
Hamiltonian in standard form ${\cal H}=\frac{p^2}{2m(x)}+{\cal V}(x)$ involves the modif\/ication
of the dynamical equations from $\dot x=\frac{\partial H_0}{\partial p}$ and $\dot p=-
\frac{\partial H_0}{\partial x}$ to those given in~\eqref{canonical}.
Hence, the simple substitution $m_0\rightarrow m(x)$ in $H_0=\frac{p^2}{2m_0}+{\cal V}(x)$ to
get~${\cal H}$ makes the problem more involved.
In the second case, the invariant $H$ is the Hamiltonian of a~variable mass $\mathrm m=m^2/m_0$
which is subjected to the ef\/fective potential ${\cal V}_{\text{ef\/f}}(x)$ rather than being acted by the
potential~${\cal V}(x)$.
In this latter picture, the dynamical laws are of the standard form (see equation~\eqref{inv6c}).
Thus, to preserve the form of the canonical equations, the mass~$m$ as well as the potential ${\cal
V}$ must be transformed into $\mathrm m$ and ${\cal V}_{\text{ef\/f}}$ respectively.
On the other hand, to preserve the form of the Hamiltonian after making $m_0\rightarrow m(x)$, the
dynamical equations must be modif\/ied according to~\eqref{canonical}.

\subsubsection{First applications}\label{first}

To close this section let us stress that, although the Lagrangian and Hamiltonian functions are not
unique, our expressions~\eqref{LH} are ensured by the conditions for the existence of standard
Lagrangians for equations with space-dependent coef\/f\/icients discussed in~\cite{Mus08}.
Let us multiply equation~\eqref{force} by $1/m(x)$, after the identif\/ication
\begin{gather*}
b(x) = \frac{m'(x)}{m(x)} =\frac{d \ln m(x)}{dx}, \qquad c(x) = \frac{1}{m(x)}, \qquad g(x) = \frac{d
{\cal V}(x)}{dx},
\end{gather*}
we arrive at
\begin{gather*}
\ddot x + b(x) \dot x^2 + c(x) g(x)=0.
\end{gather*}
So that this last equation corresponds to a~dissipative system of variable mass and admits a~Lagrangian description (see~\cite[Proposition~3]{Mus08}).
The related standard Lagrangian is
\begin{gather}
L(x, \dot x) = \frac12 \dot x^2 e^{I_b(x)} - \int^x c(\tilde x) g(\tilde x) e^{I_b(\tilde x)}
d\tilde x,
\label{qua6}
\end{gather}
where the quantity
\begin{gather*}
I_b(x) = \int^x b(\tilde x) d\tilde x = \ln m(x)
\end{gather*}
reduces~\eqref{qua6} to our Lagrangian \eqref{LH}.
This standard form of writing $L$ and ${\cal H}$ is appropriate to recover (as particular cases)
some of the Lagrangians and Hamiltonians already reported in a~dif\/ferent context by other authors.
For example, consider a~mass-function $m_{\rm e}(x)$ and a~potential function ${\cal V}(x)$ such that
\begin{gather}
m_{\rm e} (x) = m_0 e^{\kappa x/2}, \qquad \frac{d {\cal V}(x)}{d x} = e^{\kappa x/2} \frac{d V(x)}{d
x},
\label{qua1}
\end{gather}
with $m_0$ and $1/\kappa$ constants expressed in mass and position units respectively.
The Newton's equation~\eqref{force2}, the Lagrangian and the Hamiltonian \eqref{LH} become
\begin{gather*}
m_0 \ddot x = -\frac{d V(x)}{d x} - \frac12 m_0 \kappa \dot x^2,
\\
L= \frac12 m_0 \dot x^2 e^{\kappa x/2} - \int_0^x ds \, e^{\kappa s/2} \frac{d V(s)}{d s}, \qquad
{\cal H} = \frac{p^2}{2m_0} e^{-\kappa x/2} + \int_0^x ds \, e^{\kappa s/2} \frac{d V(s)}{d s}.
\end{gather*}
On the other hand, if we now take
\begin{gather*}
\underline m_{\rm e} (x)= m_0 e^{-\kappa x/4} \qquad \mathrm{and} \qquad \frac{d {\cal V} (x)}{dx} =
e^{3\kappa x/4} \frac{d V(x)}{dx}
\end{gather*}
the Hamiltonian~\eqref{hg} becomes
\begin{gather*}
H=\frac{{\mathrm p}^2}{2m_0}e^{\kappa x/2}+\int^x ds \, e^{\kappa s/2}\frac{dV(s)}{ds}.
\end{gather*}
These last expressions are in agreement with those reported in~\cite{Stu86} for a~system of
constant mass $m_0$, subjected to a~`force quadratic in the velocity' $-\frac12m_0\kappa\dot
x^2$.
An immediate generalization, considering now a~mass-function $m(x)=m_0\exp\left(\frac12\kappa(x)\right)$,
can be put in connection to the system of constant mass discussed in~\cite{Bor88}.
It is remarkable that the results for constant masses reported in~\cite{Bor88, Stu86} are also
derivable for a~mass that varies exponentially with the position, a~situation that seems to be
unnoticed in such references.
Moreover, no solutions to the corresponding canonical problem~\eqref{canonicalp} are given in
neither~\cite{Stu86} nor \cite{Bor88}.
We get explicit solutions to this problem in Section~\ref{expon}.

Next, we shall construct solutions to the system~\eqref{canonicalp} for $m(x)$ an arbitrary,
dif\/ferentiable and integrable function of~$x$, and the properly chosen potential ${\cal V}(x)$.

\subsection{Factorization and deformed algebras}\label{secfac}

The dynamical problem~\eqref{inv6c} can be studied in two general ways (compare with the quantum
problems studied in~\cite{Cru09}).
First, given a~specif\/ic potential ${\cal V}(x)$ acting on the mass $m(x)$, the related phase
trajectories are found.
Second, given an algebra which rules the dynamical law of the mass, the potential and phase
trajectories are constructed in a~purely algebraic form.
Our aim here is to follow the second approach.
For this we shall extend the factorization method discussed in~\cite{Cru08, Kur08} to the case of
classical systems having a~position-dependent mass and obeying the dynamical law~\eqref{force}.
As we have discussed in the previous sections, it is better to face the problem in the $(x,\mathrm p)$-plane.
The factorization of the Hamiltonian~\eqref{hg} leads in a~natural form to the identif\/ication of a
pair of time-dependent integrals of motion ${\cal Q}^{\pm}$ which, in turn, allows the construction
of the phase trajectories $(x(t),\mathrm p(t))$ associated to the canonical equations.

We look for a~couple of complex functions ${\cal A}^+(x,\mathrm p;t)$, ${\cal A}^-(x,\mathrm
p;t)$, and a~constant $\epsilon$ such that the Hamiltonian~\eqref{hg} becomes factorized
\begin{gather}
H=\mathcal{A}^+\mathcal{A}^-+\epsilon=\mathcal{A}^-\mathcal{A}^++\epsilon.
\label{factor1}
\end{gather}
The inspection of these last equations suggest to def\/ine ${\cal A}^{\pm}$ as follows
\begin{gather}
\mathcal{A}^{\pm}=\mp if(x)\frac{\mathrm p}{\sqrt{2\mathrm m(x)}}+g(x)\sqrt{\gamma{H}},
\label{aes}
\end{gather}
with $f(x)$ and $g(x)$ functions to be determined.
Here, we are considering the possibility of bound states (conf\/ined motion) for which the
Hamiltonian ${H}$ is negative $({H}<0)$.
Thereby, according to the sign of ${H}$, $\gamma$ will be either $+1$ or $-1$ such that
$\sqrt{\gamma{H}}$ is real in~\eqref{aes}.
The introduction of~\eqref{aes} into \eqref{factor1} leads to the relationships
\begin{gather}
f^2(x)+\gamma g^2(x)=1,
\label{setH}\\
\epsilon=\mathcal{V}_{\text{ef\/f}}(x)f^2(x).
\label{setH2}
\end{gather}
Given $\mathrm m(x)$ and $\mathcal{V}_{\text{ef\/f}}(x)$, the functions $\mathcal{A}^\pm$ and the
Hamiltonian ${H}$ induce a~Poisson structure,
\begin{gather}
i \left\{ \mathcal{A}^-,\mathcal{A}^+\right\} = \displaystyle\frac{2}{\sqrt{2 \mathrm m(x)}}
W(f,g)\sqrt{\gamma {H}},\nonumber\\
\left\{{H},\mathcal{A}^+ \right\} = {\cal A}^+ \left\{ \mathcal{A}^-,\mathcal{A}^+\right\}, \qquad
\left\{{H},\mathcal{A}^- \right\} = -\left\{ \mathcal{A}^-,\mathcal{A}^+\right\} {\cal A}^-,
\label{sys}
\end{gather}
where $W(f,g)=fg'-f'g$ is the Wronskian of $f$ and~$g$, and we have used~\eqref{setH2}.
Now, we ask the system~\eqref{sys} to close a~deformed Poisson algebra by demanding that $\{{\cal
A}^-,{\cal A}^+\}$ be expressed in terms of the powers of $\sqrt{\gamma{H}}$.
The easiest way to satisfy this condition is by looking for the solutions of equation
\begin{gather}
W(f,g)= \alpha \sqrt{2 \mathrm m(x)}, \qquad \alpha = \text{const}.
\label{algebra}
\end{gather}
Using~\eqref{setH}, this last equation is reduced to quadratures
\begin{gather}
\int\frac{dg}{\sqrt{1-\gamma g^2}}=\sqrt{2\alpha^2m_0}\int_c^x J(t)dt,\qquad
J(x)=\sqrt{\frac{\mathrm m(x)}{m_0}}.
\label{fixedfg}
\end{gather}
The constants $m_0$ and $c$ are expressed in mass and length units respectively.
Considering the sign of the Hamiltonian ${H}$, equation~\eqref{fixedfg} gives
\begin{gather}
g(x) =
\begin{cases}
\displaystyle \sin \left[ \sqrt{2 \alpha^2m_0} \int_c^x J(t)dt \right], & \gamma=1, \ \ \  {H}>0,\vspace{1mm}\\
\displaystyle \sinh \left[ \sqrt{2 \alpha^2m_0} \int_c^x J(t)dt \right], & \gamma=-1,  \  {H}<0.
\end{cases}
\label{g}
\end{gather}
The expression for $f(x)$ is obtained after the substitution of~\eqref{g} into \eqref{setH}.
At this stage we realize that the relationships~\eqref{setH} and \eqref{setH2} def\/ine the potential
${\cal V}_{\text{ef\/f}}$ in terms of the $g$-function
\begin{gather}
{\cal V}_{\text{ef\/f}} (x) = \frac{\epsilon}{1-\gamma g^2(x)} =
\begin{cases}
\displaystyle\frac{\epsilon}{\cos^2\left[ \sqrt{2 \alpha^2m_0} \int_c^x J(t)dt \right]}, & {H}>0,\\
\displaystyle\frac{\epsilon}{\cosh^2\left[ \sqrt{2 \alpha^2m_0} \int_c^x J(t)dt \right]}, & {H}<0.
\end{cases}
\label{pot}
\end{gather}
Thus, the potential allowing equations~\eqref{algebra} and \eqref{fixedfg} is not arbitrary.
As we can see, given the mass ${\mathrm m}(x)$, a~potential of the P\"oschl--Teller form~\eqref{pot}
is such that the system~\eqref{sys} becomes the deformed Poisson algebra def\/ined by
\begin{gather}
i\left\{\mathcal{A}^-,\mathcal{A}^+\right\}=2\alpha\sqrt{\gamma{H}},
\label{palgebra1}\\
i\left\{{H},\mathcal{A}^\pm\right\}=\pm2\alpha\sqrt{\gamma{H}}\mathcal{A}^\pm.
\label{palgebra2}
\end{gather}
The latter results are consistent with our factorization approach~\eqref{factor1},~\eqref{aes}.
Indeed, using relation~\eqref{palgebra2}, it is easy to verify the following Poisson brackets
\begin{gather*}
\{ { H}, {\cal A}^+ {\cal A}^- \} = \{ { H}, {\cal A}^- {\cal A}^+\} = 0.
\end{gather*}
Hence, the factorization~\eqref{factor1} makes sense because the products ${\cal A}^+{\cal A}^-$
and ${\cal A}^-{\cal A}^+$ are functions of the Hamiltonian ${H}$.
In particular, up to an arbitrary additive constant, each of them can be chosen to be proportional
to ${H}$.
Moreover, using~\eqref{palgebra2} and the Jacobi identity
\begin{gather*}
\{ { H}, \{ {\cal A}^-, {\cal A}^+ \} \} + \{ {\cal A}^-, \{ {\cal A}^+, { H} \} \} + \{ {\cal
A}^+, \{ { H}, {\cal A}^- \} \} =0,
\end{gather*}
one arrives at the Poisson bracket
\begin{gather*}
\{ { H}, \{ {\cal A}^-, {\cal A}^+ \} \} = -i 2\alpha \big\{ {\cal A}^+ {\cal A}^-, \sqrt{\gamma { H}}
\big\}=0.
\end{gather*}
So that $\{{\cal A}^-,{\cal A}^+\}$ is also a~function of ${H}$ and~\eqref{palgebra1} makes
sense.

The relevance of equation~\eqref{palgebra2} is clear, it implies both the factorization
\eqref{factor1} and the Poisson bracket (\ref{palgebra1}).
This also allows the construction of f\/irst integrals of motion for the problem we are dealing with.
Let us introduce the functions ${\cal Q}^{\pm}=\theta_{\pm}(t){\cal A}^{\pm}$, and calculate
their total time derivative.
According to~\eqref{inv7}, we have
\begin{gather}
\frac{d {\cal Q}^{\pm}}{dt} = \{ {\cal Q}^{\pm}, { H}\} + \frac{\partial {\cal Q}^{\pm}}{\partial
t} = \left[ \pm i 2 \alpha \sqrt{\gamma { H} } \theta_{\pm}(t) + \frac{\partial
\theta_{\pm}}{\partial t} \right] {\cal A}^{\pm}.
\label{invariant}
\end{gather}
The $\theta$-functions are determined by canceling the expression in square brackets.
For ${E}$ the value of~${H}$, taken as a~parameter in the context of Section~\ref{new}, one gets
$\theta_{\pm}(t)=\theta_0e^{\mp i2\alpha\sqrt{\gamma{E}}t}$ (hereafter $\theta_0=1$).
These time-dependent functions also produce a~null total time derivative~\eqref{invariant}, so that~${\cal Q}^{\pm}$ are the following time-dependent integrals of motion
\begin{gather}
\mathcal{Q}^\pm=\exp\big[\mp i2\alpha\sqrt{\gamma{E}}t\big]\mathcal{A}^\pm.
\label{inmott}
\end{gather}
Remark that $\mathcal{Q}^+$ is the complex conjugation of $\mathcal{Q}^-$, therefore
$\mathcal{Q}^+\mathcal{Q}^-=\vert\mathcal{Q}^-\vert^2={H}-\epsilon$.
Let $q^{\pm}$ be the value of the integral of motion ${\cal Q}^{\pm}$, in polar form this can be
written as
\begin{gather}
q^\pm=\sqrt{{E}-\epsilon}e^{\pm i\phi_0},\qquad\phi_0\in\mathbb R.
\label{qtrig}
\end{gather}
Let us stress that ${E}-\epsilon\geq0$ implies ${E}\geq\epsilon>0$ for ${H}>0$,
and $\vert{E}\vert\leq\vert\epsilon\vert$ for ${H}<0$.
The values of ${\cal A}^{\pm}$ can be now obtained from~\eqref{aes}, \eqref{inmott} and
\eqref{qtrig}:
\begin{gather*}
\mp i f(x) \frac{\mathrm p}{\sqrt{2 \mathrm m(x)}} + g(x) \sqrt{\gamma { E}} = \sqrt{{ E} -
\epsilon} \exp \big[\pm i (\phi_0 +2\alpha \sqrt{\gamma { E}} t)\big].
\end{gather*}
The latter equations lead to the phase trajectories $(x(t),\mathrm p(t))$ of bound states
(conf\/ined motion) with energy $E$ greater than or equal to the global minimum of the potential.
The constant $\phi_0$ must be f\/ixed by the initial conditions.
We have
\begin{gather}
g(x(t))=\displaystyle\sqrt{\frac{{E}-\epsilon}{\gamma{E}}}
\cos\big[\phi_0+2\alpha\sqrt{\gamma{E}}t\big],
\label{trayec1}\\
\mathrm p(t)=- \sqrt{\frac{2({E}-
\epsilon)}{\pi(t)}}\sin\big[\phi_0+2\alpha\sqrt{\gamma{E}}t\big],
\nonumber
\end{gather}
where
\begin{gather*}
\pi(t)= \frac{f^2(x(t))}{\mathrm m(x(t))}.
\end{gather*}
The explicit form of~$x$ as a~function of $E$ and $t$ is obtained from~\eqref{trayec1} and
\eqref{g} by using the composition $[g^{-1}\circ g](x(t))$.

\subsection{Point transformations}\label{point}

In Section~\ref{secfac} we have shown that potentials of the P\"oschl--Teller form~\eqref{pot} give
rise to the deformed Poisson algebras~\eqref{palgebra1}, \eqref{palgebra2}; these last associated to
the dynamics of a~particle of mass varying with the position.
Now, we want to make a~point transformation leading from the canonical coordinates,~$x$ and $p$, of
the variable mass~$m(x)$, to the coordinates, $Q$ and $P$, of an equivalent system of constant mass~$m_0$.
As usual, we shall write
\begin{gather*}
Q= Q(x,p;t) \qquad \text{and} \qquad P=P (x,p; t)
\end{gather*}
for the point transformation we look for.
Using~\eqref{canonical} we arrive at
\begin{gather*}
\dot Q = \{ Q, {\cal H} \} + R \frac{\partial Q}{\partial p} + \frac{\partial Q}{\partial t},
\qquad \dot P = \{ P, {\cal H} \} + R \frac{\partial P}{\partial p} + \frac{\partial P}{\partial t}.
\end{gather*}
We want these equations to have the same form as those given in~\eqref{canonical} for a~function
$K(Q,P;t)$ to be determined.
Thus, we ask for
\begin{gather}
\dot Q= \frac{\partial K}{\partial P}, \qquad \dot P = -\frac{\partial K}{\partial Q} + R_K,
\label{pec3}
\end{gather}
where $R_K$ is the reacting thrust~\eqref{Rp} expressed in the new coordinates.
For the sake of simplicity, let us assume that $Q$ is not an explicit function of neither $p$ nor
$t$, and that $P$ is time-independent.
We have
\begin{gather*}
\frac{\partial K}{\partial P} = \{ Q, {\cal H} \}, \qquad -\frac{\partial K}{\partial Q} + R_K =
\{ P, {\cal H} \} + R \frac{\partial P}{\partial p}.
\end{gather*}
Let $K$ be the Hamiltonian function associated to a~system of constant mass $m_0$ under the action
of a~given potential $V(Q)$,
\begin{gather*}
K= \frac{P^2}{2m_0} +V(Q).
\end{gather*}
One easily identif\/ies that the point transformations
\begin{gather}
Q= \int^x \sqrt{\frac{m(s)}{m_0}} ds \qquad \text{and} \qquad P=\sqrt{\frac{m_0}{m(x)}} p
\label{pec5}
\end{gather}
allow the mapping from the variable mass Hamiltonian ${\cal H}$ to the new one $K$.
It is a~matter of substitution to verify that these new variables satisfy the equations
\eqref{pec3}, explicitly
\begin{gather*}
\dot Q = \frac{P}{m_0}, \qquad \dot P = -\frac{d V(Q)}{dQ} + R_K.
\end{gather*}
That is,~\eqref{pec5} represents the canonical transformation from the dynamics of the
position-dependent mass P\"oschl--Teller potential~\eqref{pot} to the one of a~constant mass system.
The similar approach shows that the mapping $(x,\mathrm p)\rightarrow(\widetilde Q,\widetilde
P)$, with $\widetilde Q=\int^x J(s)ds$ and $\widetilde P=\mathrm p/J(x)$ leads from $H$ to the
constant mass Hamiltonian $\widetilde K=\frac{\widetilde P^2}{2m_0}+\widetilde{\cal
V}_{\text{ef\/f}}(\widetilde Q)$, where ${\cal V}_{\text{ef\/f}}(x)=\widetilde{\cal V}_{\text{ef\/f}}(\widetilde Q(x))$.

\section[Position-dependent mass P\"oschl-Teller potentials]{Position-dependent mass P\"oschl--Teller potentials}\label{PTeller}

In the previous sections we have shown that potentials of the P\"oschl--Teller form~\eqref{pot} lead
to the solving of the canonical equations of motion~\eqref{inv6c} in terms of the invariants
def\/ined by the deformed Poisson algebra~\eqref{palgebra1}, \eqref{palgebra2}.
Depending on the sign of the Hamiltonian~${H}$, equation~\eqref{pot} includes two general forms of
these potentials and, as a~consequence, we also have two dif\/ferent realizations of the algebra~\eqref{palgebra1}, \eqref{palgebra2}.
Namely, for a~positive Hamiltonian ${H}>0$, the identif\/ication
\begin{gather*}
a^{\pm} = \frac{1}{\alpha} {\cal A}^{\pm}, \qquad a^0 = \frac{1}{\alpha} \sqrt{ { H}},
\end{gather*}
makes clear that the Poisson structure~\eqref{palgebra1}, \eqref{palgebra2} associated to the
dynamical algebra of the trigonometric position-dependent mass P\"oschl--Teller potential
\begin{gather*}
{\cal V}_{\text{ef\/f}}(x) =\frac{\epsilon}{\cos^2\left[ \sqrt{2 \alpha^2m_0} \int_c^x J(t)dt \right]}
\end{gather*}
corresponds to the $su(1,1)$ algebra
\begin{gather*}
i\left\{a^-,a^+\right\}=2a^0,\qquad i\left\{a^0,a^\pm\right\}=\pm a^\pm.
\end{gather*}
In a~similar form, for a~negative Hamiltonian ${H}<0$, the identif\/ication
\begin{gather*}
a^\pm=\frac{1}{\alpha}\mathcal{A}^\pm,\qquad a^0=-\frac{1}{\alpha}\sqrt{-{H}},
\end{gather*}
shows that the dynamics of a~position-dependent mass particle subjected to the hyperbolic
P\"oschl--Teller potential
\begin{gather*}
{\cal V}_{\text{ef\/f}}(x) =\frac{\epsilon}{\cosh^2\left[ \sqrt{2 \alpha^2m_0} \int_c^x J(t)dt \right]}
\end{gather*}
is nothing but the $su(2)$ Poisson algebra
\begin{gather*}
i\left\{a^-,a^+\right\}=-2a^0,\qquad i\left\{a^0,a^\pm\right\}=\pm a^\pm.
\end{gather*}
It \looseness=-1  is well known that the above algebras are related to the P\"oschl--Teller systems of constant
mass.
Here, we have shown that this is the case even if the mass is a~function of the position.
The set of P\"oschl--Teller potentials is relevant because of its diversity of applications.
For instance, in the constant mass quantum case such a~set has been enlarged to a~wide family of
supersymmetric potentials~\cite{Cont08, Dia99}.
Another example can be found in the study of gravity localization and thick branes of cosmological
problems~\cite{Guo12a,Guo12b}.
Approaches similar to the one presented here have been also applied to get a~subalgebra of the
Poisson algebra in the constant mass case~\cite{Gue09}.
There, one can f\/ind transformations mapping the whole phase space of the trigonometric case into a~whole coadjoint orbit of $SO(2,1)$.
The hyperbolic case deserves some caution because, even in the constant mass case, the positive and
negative energies give rise to dif\/ferent Lie groups.

Next, we are going to discuss some specif\/ic realizations of our general results by supplying
dif\/ferent forms of the mass-function $\mathrm m(x)$.
In each case, the phase trajectories $(x(t),\mathrm p(t))$ are explicitly derived.

\subsection{Doubly singular mass-functions}
\label{doubly}

The mass function
\begin{gather}\label{mass0}
m_{\rm ss}(x)=m_0\left(\frac{1-\beta(\lambda x)^2}{1-(\lambda x)^2}\right),
\qquad\lambda>0,\qquad\beta<0,
\end{gather}
has been used in the study of the inversion potential for NH$_3$ in terms of the density theory~\cite{Aqu98}.
This function is singular at the points $x=\pm1/\lambda$, and has no zeros in the interval
$\left(-1/\lambda,1/\lambda\right)$.
In notation of~\cite{Aqu98}, $m_0=\mu_0$ should correspond to the reduced mass
\begin{gather*}
\mu_0=\frac{3mM}{3m+M},
\end{gather*}
with~$m$ the mass of each of the three hydrogen atoms, and~$M$ the mass of the nitrogen atom in the
geometry of a~rectangular pyramid where~$M$ is in the cusp.
The mass-function~\eqref{mass0} then corresponds to the situation in which the distance between
each pair of masses~$m$ is allowed to change.
The parameter $1/\lambda=r_0$ stands for the separation between~$m$ and~$M$ at the planar
equilibrium geometry,~$x$ is the inversion coordinate and the negative parameter $\beta$ is given
by $\beta=-3m/M$.
Here, $m_0>0$ and $1/\lambda>0$ are arbitrary constants expressed in mass and distance units
respectively.
The negative, dimensionless parameter $\beta$ is also arbitrary.

\begin{figure}[t]\centering
\includegraphics[width=14.5cm]{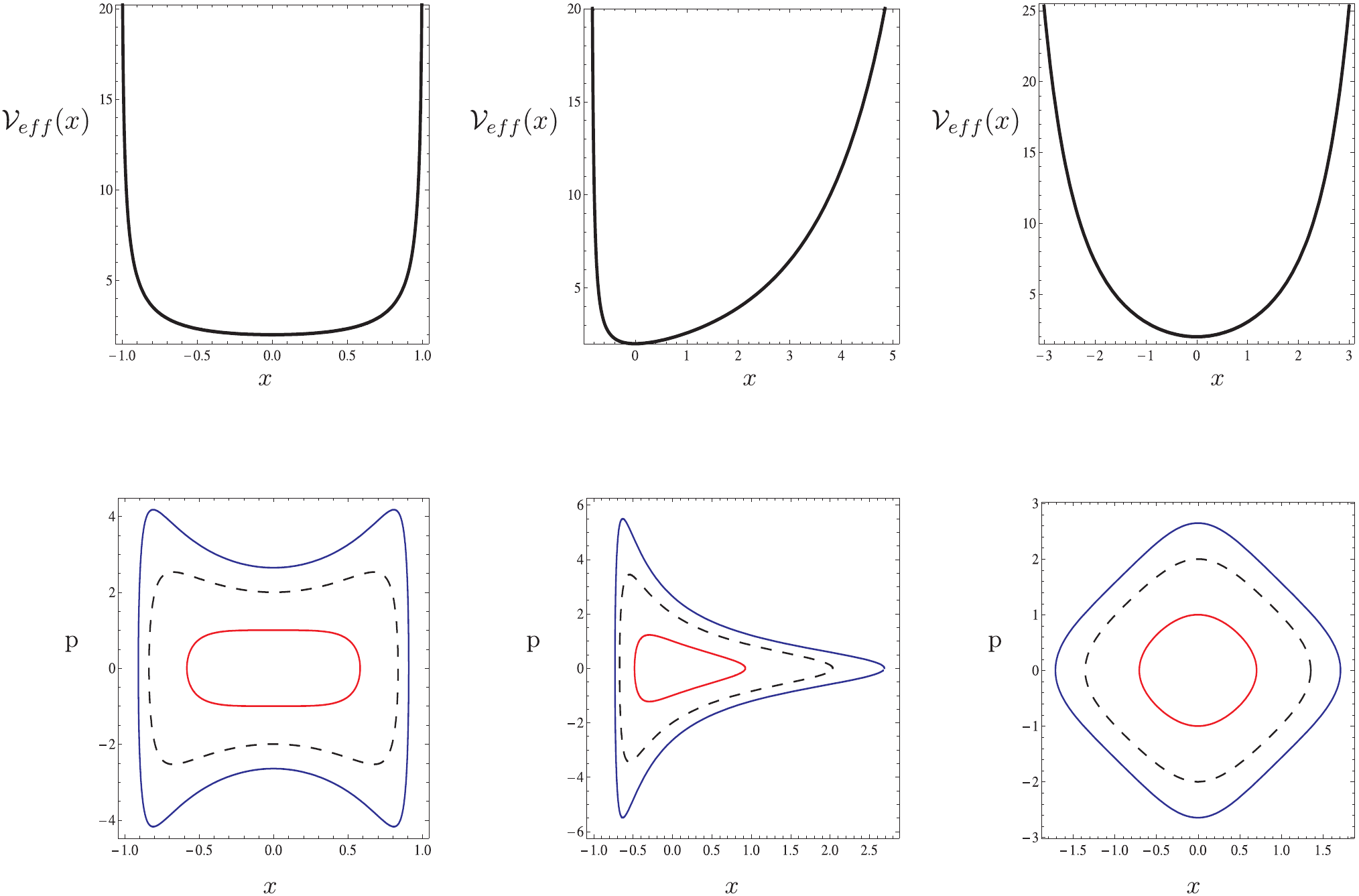}

\caption{\textit{Upper line.} Position dependent-mass trigonometric P\"oschl--Teller potentials for the
mass-func\-tions $m_{\rm ss}$ (left), $m_{\rm s}$ (center), $m_R$ (right), and the parameters $\alpha=1/2$,
$\beta=-1$, $m_0=\lambda=1$.
{\it Lower line.} Some of the related (conf\/ined motion) classical trajectories in phase space for
$\phi_0=0$, $\epsilon=2$ and ${E}=2.5$ (red curve), ${E}=4$ (dashed curve), ${E}=5.5$ (blue
curve).}\label{fig1}
\end{figure}

\begin{figure}[t]\centering
\includegraphics[width=14.5cm]{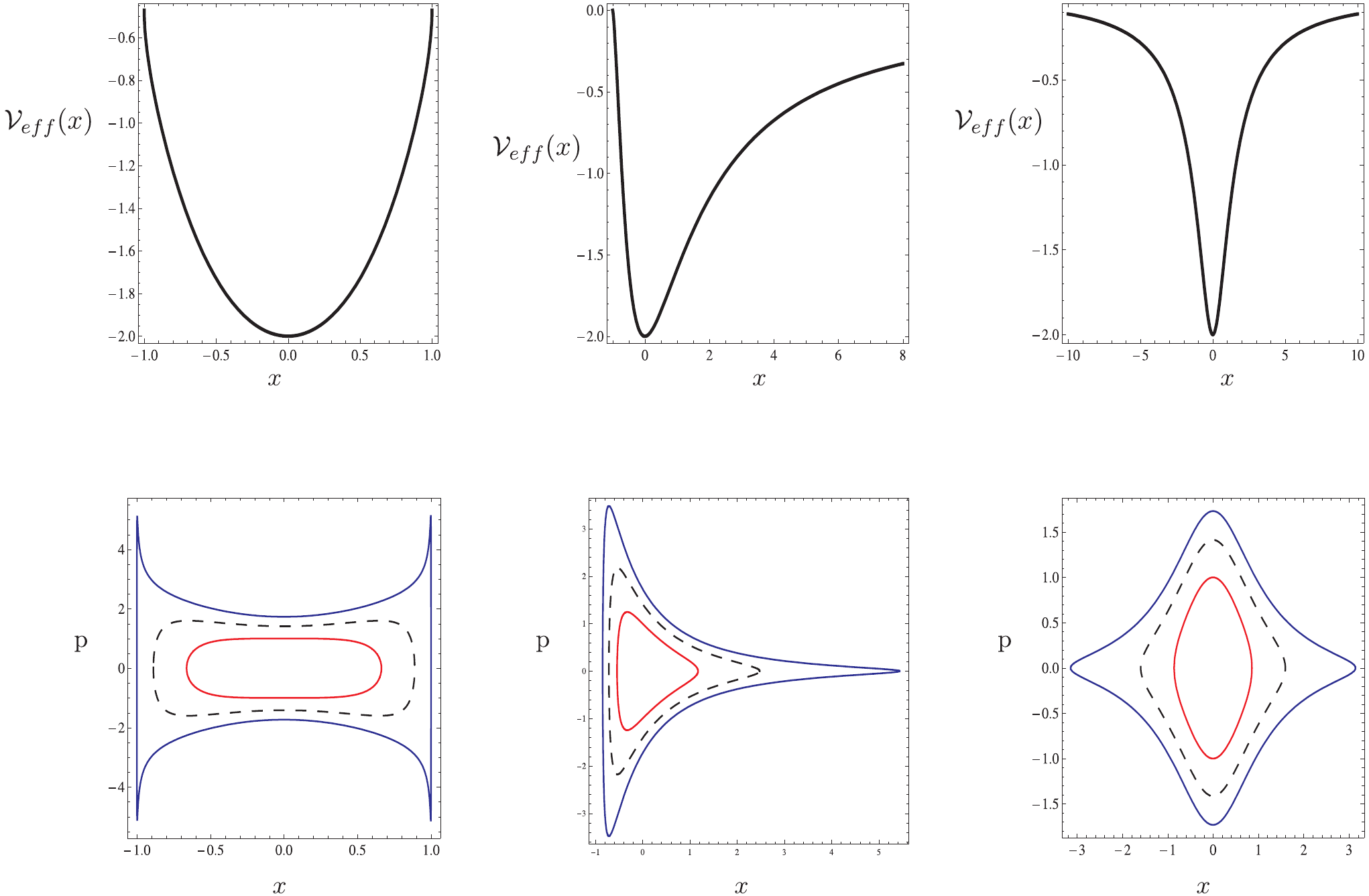}

\caption{{\it Upper line.} Position dependent-mass hyperbolic P\"oschl--Teller potentials for the
mass-functions $m_{\rm ss}$ (left), $m_{\rm s}$ (center), $m_R$ (right), and the parameters $\alpha=1/2$,
$\beta=-1$, $m_0=\lambda=1$.
{\it Lower line.} Some of the related (conf\/ined motion) classical trajectories in phase space for
$\phi_0=0$, $\epsilon=-2$ and ${E}=-1.5$ (red curve), ${E}=-1$ (dashed curve), ${E}=-0.5$
(blue curve).}
\label{fig2}
\end{figure}

To get explicit forms of the solutions we use~\eqref{fixedfg} and arrive at
\begin{gather*}
\int_0^x J_{\rm ss}(t)dt=\int_0^x\sqrt{\frac{m_{\rm ss}(t)}{m_0}}dt=
\frac{1}{\lambda}\int_0^{\lambda x}\sqrt{\frac{1-\beta t^2}{1-t^2}}dt=\frac{1}{\lambda}
E_{\rm int}\big(\arcsin(\lambda x),\beta^{1/2}\big)
\end{gather*}
(hereafter we take $c=0$), with $E_{\rm int}(\phi,k)$ the incomplete elliptic integral of the second
kind~\cite{Gra94}
\begin{gather*}
E_{\rm int}\left(\phi,k\right)=\int_0^{\sin\phi}\sqrt{\frac{1-(k t)^2}{1-t^2}}dt.
\end{gather*}
Potentials~\eqref{pot} are in this case given by
\begin{gather*}
{\cal V}_{\rm ss}(x)=
\begin{cases}
\displaystyle\frac{\epsilon}{\cos^2 \left[\sqrt{2m_0} \frac{\alpha}{\lambda} E_{\rm int} \big(\arcsin
\lambda x,\beta^{1/2}\big)\right]}, & {H}>0,\vspace{1mm}\\
\displaystyle\frac{\epsilon}{\cosh^2 \left[\sqrt{2m_0} \frac{\alpha}{\lambda} E_{\rm int}
\left(\arcsin\lambda x,\beta^{1/2} \right)\right]}, & {H}<0.
\end{cases}
\end{gather*}
The domains of def\/inition for these potentials are
\begin{gather*}
{\cal D}_{{\rm ss}>}=\left\{x\left\vert-\frac{\pi}{2}<
\sqrt{2m_0}\frac{\alpha}{\lambda}E_{\rm int}\big(\arcsin\lambda x,\beta^{1/2}\big)
<\frac{\pi}{2}\right.\right\},
\end{gather*}
and ${\cal D}_{{\rm ss}<}=(-1/\lambda,1/\lambda)$ respectively.
Finally, the corresponding phase trajectories are obtained from the expressions
\begin{gather*}
 \sin \sqrt{2m_0}\frac{\alpha}{\lambda} E_{\rm int}\big(\arcsin \lambda
x(t),\beta^{1/2}\big) = \sqrt{\frac{{E}-\epsilon}{{E}}} \cos \big(2\sqrt{{E}}\alpha t +
\phi_0\big),\\
  \mathrm p(t) = -\frac{\sqrt{2m_0({E}-\epsilon)\left(1-\beta(\lambda
x(t))^2\right)}}{\sqrt{1-\left(\lambda x(t)\right)^2}\cos
\left[\sqrt{2m_0}\frac{\alpha}{\lambda}E_{\rm int}\left(\arcsin \lambda
x(t),\beta^{1/2}\right)\right]}\sin \big(2\sqrt{{E}}\alpha t +\phi_0\big),
\end{gather*}
and
\begin{gather*}
  \sinh \sqrt{2m_0}\frac{\alpha}{\lambda} E_{\rm int}\big(\arcsin \lambda x(t),
\beta^{1/2}\big) = \sqrt{\frac{{E}-\epsilon}{-{E}}}\cos \big(2\sqrt{-{E}}\alpha t
+\phi_0\big),\\
\mathrm p(t) =- \frac{\sqrt{2m_0({E}-\epsilon)\left(1-\beta(\lambda
x(t))^2\right)}}{\sqrt{1-\left(\lambda x(t)\right)^2}\cosh
\left[\sqrt{2m_0}\frac{\alpha}{\lambda}E_{\rm int} \left(\arcsin \lambda
x(t),\beta^{1/2}\right)\right]}\sin \big(2\sqrt{-{E}}\alpha t +\phi_0\big).
\end{gather*}
The potentials ${\cal V}_{\rm ss}(x)$ and phase trajectories $(x(t),\mathrm p(t))$ have been depicted
in Figs.~\ref{fig1} and~\ref{fig2} for dif\/ferent values of the parameters.
From Fig.~\ref{fig1} (left) we realize that only conf\/ined motion is allowed for the domain ${\cal D}_{{\rm ss}>}$.
The mass $m_{\rm ss}(x)$ is singular at the edges of ${\cal D}_{{\rm ss}>}$, where the particle reaches the
turning points and its momentum $\mathrm p$ is zero.
The minimal amount of mass is obtained at origin, so that the mass increases in value as the
particle approaches the turning points and conversely, the particle losses mass as it approaches
the origin.
The momentum $\mathrm p$ evolves in time in a~quite similar manner: For bounded energies greater
than the global minimum of the potential, the particle acquires a~momentum $\mathrm p(t)$ which is
greater as the particle approaches one of the turning points.
Once there, the momentum of the particle changes in sign so that the motion is reverted by a~strong acceleration.
This relationship between the maxima of the mass and the maxima of the momentum is also found in
the conf\/ined motion for $H<0$ (Fig.~\ref{fig2}, left).
There, it is also true that the momentum $\mathrm p$ is zero at the points where the mass is divergent.
The same can be said for the other masses analyzed in the sequel.
Of particular interest, the regular mass case (discussed in Section~\ref{sregular}) is such that
the momentum~$\mathrm p$ and the mass-function~$m_R(x)$ are maxima at the origin, and both of them
take their lower admissible values at the corresponding turning points.

\subsection{Singular mass-functions}

Given a~mass function $m(x)$, the quantum problem for the potential $V(x)$ can be solved by the
mapping of the Schr\"odinger equation of $m(x)$ to the Schr\"odinger equation of a~constant mass~$m_0$ (for a~general discussion see~\cite{Cru09}).
In the trivial case of a~constant mass~$m_0$, the new poten\-tial~$V_{\text{ef\/f}}$ (expressed in the new
coordinates) is the same as the former one $V$.
In general, the mass function
\begin{gather}
m_{\rm s}(x) = \frac{m_0}{(1+\lambda x)^2}
\label{mass1}
\end{gather}
has been shown to be the simplest nontrivial case in which $V_{\text{ef\/f}}=V$~\cite{Cru09}.
This mass is also connected with the revival of wave packets in a~position-dependent mass inf\/inite
well~\cite{Sch06}, and was used in our previous analysis of diverse position-dependent mass
oscillators~\cite{Cru07,Cru08b,Cru09,Cru10}.
The function~\eqref{mass1} is singular at the point $x=-1/\lambda$ and has no zeros in
$(-1/\lambda,+\infty)$.
The potentials~\eqref{pot} read as
\begin{gather*}
{\cal V}_{\rm s}(x) =
\begin{cases}
\displaystyle\frac{\epsilon}{\cos^2\left[\sqrt{2m_0}\frac{\alpha}{\lambda}\ln \left(1+\lambda
x\right)\right]}, & {H}>0,\vspace{1mm}\\
\displaystyle\frac{\epsilon}{\cosh^2\left[\sqrt{2m_0}\frac{\alpha}{\lambda}\ln \left(1+\lambda
x\right)\right]}, & {H}<0,
\end{cases}
\end{gather*}
with the domains of def\/inition
\begin{gather*}
\mathcal{D}_{{\rm s}>} =\left\{x \left\vert -\frac{\pi}2< \sqrt{2m_0}\frac{\alpha}{\lambda}\ln
\left(1+\lambda x\right) < \frac{\pi}2 \right.
\right\},
\end{gather*}
and $\mathcal{D}_{{\rm s}<}=\left[-1/\lambda,\infty\right)$ respectively.
The corresponding phase trajectories are ruled by the expressions
\begin{gather*}
x(t)=
\begin{cases}
\displaystyle\frac{1}{\lambda} \left[ \exp \left( \frac{\lambda}{\alpha \sqrt{2m_0}}
\arcsin \left[ \sqrt{\frac{{E}- \epsilon}{{E}}}\cos \big(2\sqrt{{E}}\alpha t + \phi_0\big)
\right]\right) -1\right], & {H}>0,\vspace{1mm}\\
\displaystyle \frac1{\lambda} \left[ \exp \left ( \frac{\lambda}{\alpha \sqrt{2m_0}}
\mathrm{arcsinh} \left[\sqrt{\frac{{E}- \epsilon}{-{E}}}
\cos \big(2\sqrt{-{E}}\alpha t + \phi_0\big)\right]\right) -1 \right], & {H}<0,
\end{cases}
\end{gather*}
and
\begin{gather*}
\mathrm p(t)=
\begin{cases}
-\displaystyle \frac{\sqrt{2m_0\left({E}-\epsilon\right)}\sin\big(2\sqrt{{E}}\alpha t
+\phi_0\big)}{\left(1 + \lambda x(t)\right)\cos \left[\sqrt{2m_0}\frac{\alpha}{\lambda}\ln
\left(1+\lambda x(t)\right)\right]}, & {H}>0,\vspace{1mm}\\
-\displaystyle\frac{\sqrt{2m_0\left({E}-\epsilon\right)}\sin\big(2\sqrt{-{E}}\alpha t
+\phi_0\big)}{\left(1 + \lambda x(t)\right)\cosh \left[\sqrt{2m_0}\frac{\alpha}{\lambda}\ln
\left(1+\lambda x(t)\right)\right]}, & {H}<0.
\end{cases}
\end{gather*}
The potentials ${\cal V}_{\rm s}(x)$ and phase trajectories $(x(t),\mathrm p(t))$ have been depicted
in Figs.~\ref{fig1} and~\ref{fig2} for dif\/ferent values of the parameters.
The description of the behavior of $\mathrm p(t)$ and $\mathrm m(x)$ is similar to that given in
Section~\ref{doubly}.

\subsection{Regular mass-functions}\label{sregular}

The mass function
\begin{gather*}
m_R(x) = \frac{m_0}{1+(\lambda x)^2}
\end{gather*}
appears in the study of diverse oscillators including the nonlinear one~\cite{Car07, Car04}, and the
related position-dependent mass versions~\cite{Cru07,Cru08b,Cru09}.
This is a~regular function def\/ined on the whole real line.
The P\"oschl--Teller potentials
\begin{gather*}
{\cal V}_R(x) =
\begin{cases}
\displaystyle\frac{\epsilon}{\cos^2 \left[\sqrt{2m_0}\frac{\alpha}{\lambda} \operatorname{arcsinh}
\lambda x\right]}, & {H}>0,\vspace{1mm}\\
\displaystyle\frac{\epsilon}{\cosh^2 \left[\sqrt{2m_0}\frac{\alpha}{\lambda} \operatorname{arcsinh}
\lambda x \right]}, & {H}<0
\end{cases}
\end{gather*}
are def\/ined in
\begin{gather*}
\mathcal{D}_{R>}= \left\{x \left\vert -\frac{\pi}2< \sqrt{2m_0}\frac{\alpha}{\lambda}
\operatorname{arcsinh} \lambda x < \frac{\pi}2 \right.
\right\},
\end{gather*}
and ${\cal D}_{R<}=\mathbb R$ respectively.
The phase trajectories read as
\begin{gather*}
x(t) =
\begin{cases}
\displaystyle\frac1{\lambda}\sinh \left[\frac{\lambda}{\alpha \sqrt{2m_0}}
\arcsin \left(\sqrt{\frac{{E}-\epsilon}{{E}}}
\cos\big(2 \sqrt{{E}} \alpha t + \phi_0\big)\right)\right], & {H}>0,\vspace{1mm}\\
\displaystyle\frac1{\lambda}\sinh \left[\frac{\lambda}{\alpha \sqrt{2m_0}}
\operatorname{arcsinh} \left(\sqrt{\frac{{E}-\epsilon}{-{E}}}
\cos\big(2\sqrt{-{E}} \alpha t + \phi_0\big)\right)\right], & {H}>0,
\end{cases}
\end{gather*}
and
\begin{gather*}
\mathrm p(t) =
\begin{cases}
-\displaystyle\sqrt{\frac{2m_0\left({E}-\epsilon\right)}{1+\left(\lambda x(t)\right)^2}}
\frac{\sin\big( 2 \sqrt{{E}}\alpha t + \phi_0\big)}{\cos
\left[\frac{\alpha}{\lambda}\sqrt{2m_0}\operatorname{arcsinh} \left(\lambda x(t)\right)\right]}, & {H}>0, \vspace{1mm}\\
-\displaystyle\sqrt{\frac{2m_0\left({E}-\epsilon\right)}{1+\left(\lambda x(t)\right)^2}}
\frac{\sin\big(2\sqrt{-{E}}\alpha t + \phi_0\big)}{\cosh\left[
\frac{\alpha}{\lambda}\sqrt{2m_0}\operatorname{arcsinh} \left(\lambda x(t)\right)\right]}, & {H}<0.
\end{cases}
\end{gather*}
The potentials ${\cal V}_{R}(x)$ and phase trajectories $(x(t),\mathrm p(t))$ have been depicted
in Figs.~\ref{fig1} and~\ref{fig2} for dif\/ferent values of the parameters.
The description of the behavior of~$\mathrm p(t)$ and~$\mathrm m(x)$ is similar to that given in
Section~\ref{doubly}.

\subsection{Exponential mass-functions}\label{expon}

\begin{figure}[t]\centering
\includegraphics[width=14.5cm]{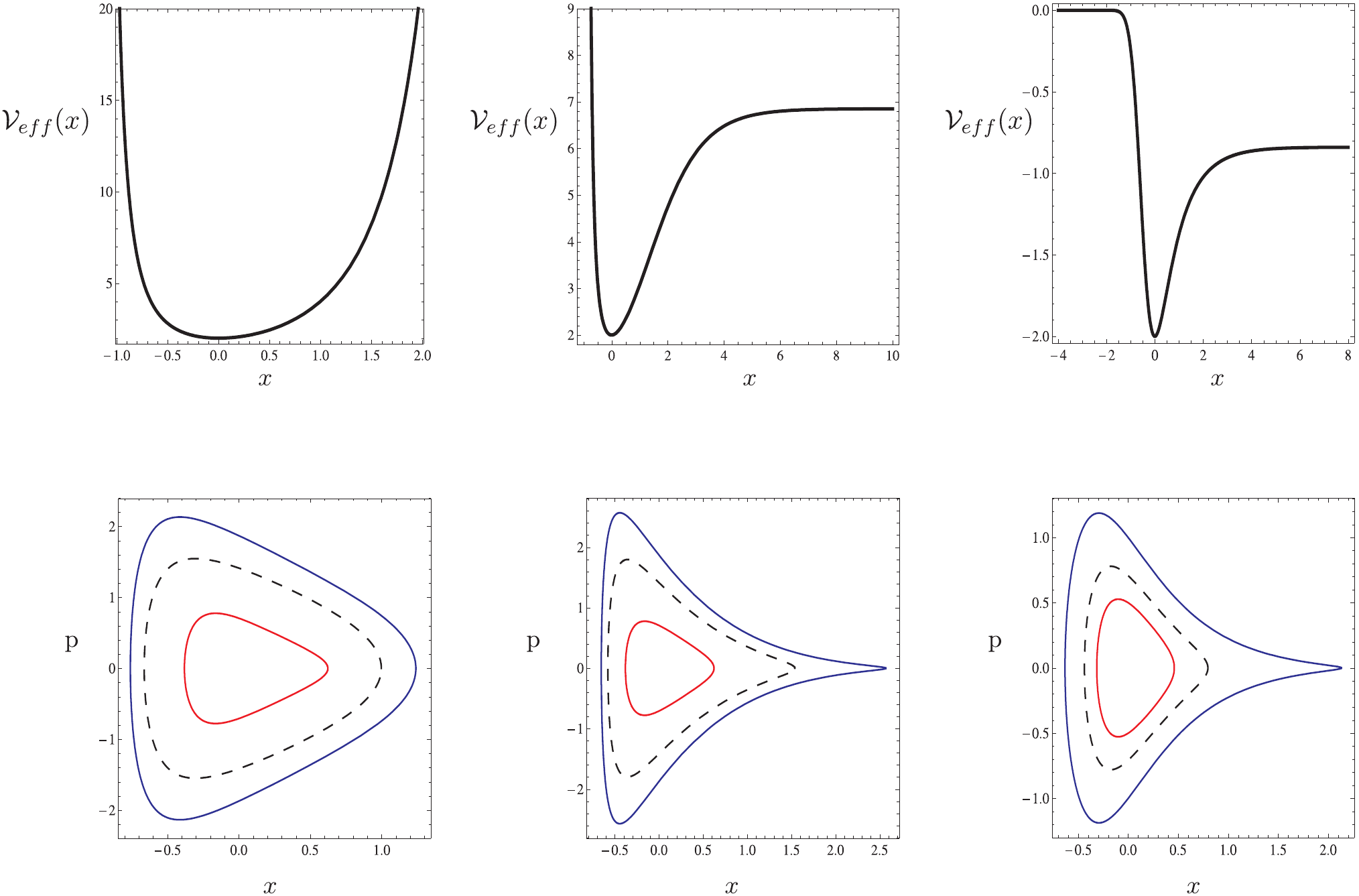}

\caption{{\it Upper line.} Position dependent-mass P\"oschl--Teller potentials for the exponential
mass-function $m_{\rm e}(x)$ in the domains of def\/inition ${\cal D}_{{\rm e}>}^1$ (left), ${\cal D}_{{\rm e}>}^2$
(center), and ${\cal D}_{{\rm e} <}$ (right).
The parameters are $\alpha=1$, $m_0=1/2$, and $\epsilon=2$, $\kappa=-2$ (left), $\epsilon=2$,
$\kappa=-4$ (center), $\epsilon=-2$, $\kappa=-4$ (right).
{\it Lower line.} Some of the related (conf\/ined motion) classical trajectories in phase space for
$\phi_0=0$ and ${E}=2.5,4,5.5$ (left and center), ${E}=-1.75,-1.5,-1$ (right).}
\label{fig3}
\end{figure}

The exponential mass-function $m_{\rm e} (x)$, introduced in equation~\eqref{qua1}, is a~regular function
def\/ined on the whole real line.
Here, we shall take $\kappa<0$ in order to get f\/inite masses in the positive regime of the domain.
The domains of def\/inition of the P\"oschl--Teller potentials
\begin{gather}
{\cal V}_{\rm e} (x)=
\begin{cases}
\displaystyle\frac{\epsilon}{\cos^2 \left[ \sqrt{2 \alpha^2 m_0} \frac{4}{\kappa} \left( e^{\kappa
x/4} -1\right) \right] }, & {H}>0,\vspace{1mm} \\
\displaystyle\frac{\epsilon}{\cosh^2 \left[ \sqrt{2 \alpha^2 m_0} \frac{4}{\kappa} \left( e^{\kappa
x/4} -1\right) \right] }, & {H}<0
\end{cases}
\label{pote}
\end{gather}
are respectively given by
\begin{gather*}
\begin{cases}
{\cal D}_{{\rm e}>}^1 = [x_0^-, x_0^+], & \text{for any} \  \kappa<0, \ m_0>0, \ \mathbb R \ni \alpha \neq
0,\\
{\cal D}_{{\rm e} >}^2= [x_0^-, +\infty), & \text{with} \  8\sqrt{2 \alpha^2 m_0} \leq \vert \kappa
\vert \pi,
\end{cases}
\end{gather*}
and ${\cal D}_{{\rm e}<}=\mathbb R$, with
\begin{gather*}
x_0 ^{\pm}= -\frac{4}{\vert \kappa \vert} \ln \left[1 \mp \frac{\pi \vert \kappa \vert}{8\sqrt{2
\alpha^2 m_0}} \right].
\end{gather*}
Potential ${\cal V}_{\rm e}(x)$ in~\eqref{pote} is singular at $x=x_0^{\pm}$ for the domain ${\cal
D}_{{\rm e}>}^1$.
This is also singular at $x=x_0^{\pm}$, and acquires a~f\/inite value as $x\rightarrow+\infty$ in
the domain ${\cal D}_{{\rm e}>}^2$.
Finally, this potential cancels as $x\rightarrow-\infty$, and goes to a~f\/inite value as $x
\rightarrow+\infty$ in the domain ${\cal D}_{{\rm e}<}$ (see Fig.~\ref{fig3}).
The related phase trajectories are constructed according to
\begin{gather}
x(t) =
\begin{cases}
\displaystyle\frac{4}{\kappa} \ln \left[ \frac{\kappa}{4\sqrt{2\alpha^2 m_0}} \arcsin \left(
\sqrt{\frac{{ E} -\epsilon}{ E} }\cos\big(\phi_0 + 2\alpha \sqrt{E} t\big) \right) + 1\right], & {H}>0,\vspace{1mm} \\
\displaystyle\frac{4}{\kappa} \ln \left[ \frac{\kappa}{4\sqrt{2\alpha^2 m_0}} \operatorname{arcsinh}
\left(\sqrt{\frac{{ E} -\epsilon}{- E} }\cos\big(\phi_0 + 2\alpha \sqrt{- E} t\big)\right) + 1\right], & {H}<0,
\end{cases}\label{xe}
\end{gather}
and
\begin{gather}
\mathrm p(t) =
\begin{cases}
-\displaystyle\frac{\sqrt{2({ E} -\epsilon) m_0} e^{\kappa x/4} \sin\big( \phi_0 + 2\alpha \sqrt{E}
t\big) }{\cos \left[ \sqrt{2 \alpha^2 m_0} \frac{4}{\kappa} \left( e^{\kappa x/4} -1 \right)
\right]}, & {H}>0,\vspace{1mm}\\
-\displaystyle\frac{\sqrt{2({ E} -\epsilon) m_0} e^{\kappa x/4} \sin\big( \phi_0 + 2\alpha \sqrt{-
E} t \big)}{\cosh \left[ \sqrt{2 \alpha^2 m_0} \frac{4}{\kappa} \left( e^{\kappa x/4} -1 \right)
\right]}, & {H}<0,
\end{cases}
\label{pe}
\end{gather}
The potentials ${\cal V}_{\rm e}(x)$ and phase trajectories $(x(t),\mathrm p(t))$ have been depicted in
Fig.~\ref{fig3} for dif\/ferent values of the parameters.
In this case we can distinguish three general kinds of potential: I.~The potential is such that
only conf\/ined motion is allowed (Fig.~\ref{fig3}, left).
II.~The potential ${\cal V}_{\rm e}(x)$ is such that scattering and positive bounded states (conf\/ined motion)
are allowed (Fig.~\ref{fig3}, center).
III.~Negative and positive energies can be associated to scattering states, and negative bounded states
are allowed (Fig.~\ref{fig3}, right).
The description of the behavior of $\mathrm p(t)$ and $\mathrm m(x)$ is similar to that given in
Section~\ref{doubly}.
With the trajectories in the phase space~\eqref{xe} and \eqref{pe} we provide an explicit solution
to the canonical problem involving ``forces quadratic in the velocity'' discussed in~\cite{Stu86} and~\cite{Bor88}.

\section{Conclusions}

\looseness=-1
We have constructed the Lagrangian for a~particle suf\/fering a~spatial variation of its mass.
The corresponding Euler--Lagrange equations have been shown to recover the Newton's dynamical law
associated to this system.
As a~consequence of the position-dependence of the mass, there is a~force quadratic in the velocity
which is non-inertial and represents the thrust of the system.
The Lagrangian for this dissipative system is of the standard form $L=\frac{p^2}{2m(x)}-{\cal
V}(x)$, in correspondence with the conditions studied in~\cite{Mus08}.
The construction of the related Hamiltonian also leads to the standard form ${\cal H}=
\frac{p^2}{2m(x)}+{\cal V}(x)$.
Thus, we have found that a~simple description of this system starts by the replacing of the
(constant) mass~$m_0$ by the appropriate function of the position $m(x)$ in the conventional
expressions of~$L$ and ${\cal H}$.
Accordingly, it has been shown that the canonical equations of motion must contain a~term including
the thrust in order to recover the Newton's equation.
Since ${\cal H}$ is not time-independent, we have explicitly constructed an energy constant of
motion $H$ leading to dynamical equations that have the form of the Hamilton ones.
A canonical transformation mapping the variable mass problem to the one of a~constant mass has been
also identif\/ied.
Departing from the factorization of the Hamiltonian $H={\cal A}^+{\cal A}^-+\epsilon$, we have
obtained two time-dependent integrals of motion ${\cal Q}^{\pm}$ which, in turn, allow the
construction of the trajectories in the phase space $(x(t),\mathrm p(t))$.
Such invariants are associated to potential functions which are necessarily of the P\"oschl--Teller
form if one looks for the related spectrum generating algebras.
The latter are obtained from the Poisson structure def\/ined by the factorization of the Hamiltonian
and demanding that the Poisson brackets of ${\cal A}^{\pm}$ and $H$ be expressed in terms of the
polynomials of $H$.
Two general forms of the P\"oschl--Teller potentials have been found, one is of the trigonometric
type and is connected with the $su(1,1)$ Lie algebra, the other is of hyperbolic type and is
associated to the $su(2)$ algebra.
Dif\/ferent solutions for the canonical equations of motion have been explicitly given in terms of
specif\/ic forms of the mass-function~$\mathrm m(x)$.
As particular cases, the Lagrangians and Hamiltonians already reported by other authors~\cite{Bor88, Stu86} have been recovered by the appropriate selection of the mass function~$m(x)$.
Moreover, in contradistinction with such works, we provide explicit solutions to the corresponding
equations of motion.
We stress that the singular oscillator, generalized P\"oschl--Teller and the Morse potentials can be
also included in our approach but they require factorizing functions ${\cal A}^{\pm}$ dif\/ferent
from the ones used in the present work.
Indeed, the underlying Poisson structures of these last systems do not always give rise to a~Lie
algebra (see for instance~\cite{Kur08}).
Finally, we mention that similar approaches to that presented here have been applied in the
analysis of the constant mass versions of the P\"oschl--Teller and Morse potentials for classical
and quantum dynamics~\cite{Ald05, Gue09}.

\subsection*{Acknowledgements}

The authors thank the anonymous referees for their comments to improve the presentation and
motivation of the paper.
The f\/inancial support of CONACyT-Mexico (project 152574), MICINN-Spain (project MTM2009-10751),
IPN grant COFAA and projects SIP20120451, SIP-SNIC-2011/04, is acknowledged.

\pdfbookmark[1]{References}{ref}
\LastPageEnding

\end{document}